\documentclass[english]{article}
\usepackage[T1]{fontenc}
\usepackage[latin9]{inputenc}
\usepackage{color}
\usepackage{amstext}
\usepackage{amssymb}
\usepackage{graphicx}
\usepackage{wasysym}
\makeatletter
\newcommand{\lyxaddress}[1]{
\par {\raggedright #1
\vspace{1.4em}
\noindent\par}
}

\usepackage{pifont}

\usepackage{babel}
\begin{document}

\title{Emergence of low noise \emph{frustrated} states in E/I balanced neural
networks}

\author{I. Recio and J.J. Torres}

\maketitle

\lyxaddress{Department of Electromagnetism and Physics of the Matter and Institute
``Carlos I'' for Theoretical and Computational Physics, University
of Granada, Granada, Spain, E-18071.}
\begin{abstract}
We study emerging phenomena in binary neural networks where, with
a probability $c$ synaptic intensities are chosen according with
a Hebbian prescription, and with probability $(1-c)$ there is an
extra random contribution to synaptic weights. This new term, randomly
taken from a Gaussian bimodal distribution, balances the synaptic
population in the network so that one has $80\%-20\%$ relation in
E/I population ratio, mimicking the balance observed in mammals cortex.
For some regions of the relevant parameters, our system depicts standard
memory (at low temperature) and non-memory attractors (at high temperature).
However, as $c$ decreases and the level of the underlying noise also
decreases below a certain temperature $T_{t}$, a kind of memory-frustrated
state, which resembles spin-glass behavior, sharply emerges. Contrary
to what occurs in Hopfield-like neural networks, the frustrated state
appears here even in the limit of the loading parameter $\alpha\rightarrow0$.
Moreover, we observed that the frustrated state in fact corresponds
to two states of non-vanishing activity uncorrelated with stored memories,
associated, respectively, to a high activity or Up state and to a
low activity or Down state. Using a linear stability analysis, we
found regions in the space of relevant parameters for locally stable
steady states and demonstrated that frustrated states coexist with
memory attractors below $T_{t}$. Then, multistability between memory
and frustrated states is present for relatively small $c,$ and metastability
of memory attractors can emerge as $c$ decreases even more. We studied
our system using standard mean-field techniques and with Monte Carlo
simulations, obtaining a perfect agreement between theory and simulations.
Our study can be useful to explain the role of synapse heterogeneity
on the emergence of stable Up and Down states not associated to memory
attractors, and to explore the conditions to induce transitions among
them, as in sleep-wake transitions. 

\end{abstract}

\section{Introduction}

Traditional Hopfield-like networks \cite{hopfield}, which assume
a Hebbian learning rule \cite{hebb,tsodykshebb} for synaptic intensities,
have been widely tested to be convenient for recall of learned memories.
In these networks, one can train the system to learn some predefined
patterns of neural activity (e. g., related with some sensory information)
through their storage at the synaptic intensities. After the learning
process, and due to the synaptic modification it implies, each neuron
in the network can be more or less excitable according with the strength
of the synaptic intensities it receives. In this way, a Hopfield-like
network is able to retrieve one of these stored patterns if the network
receives an input similar to it by means the \emph{associative memory
mechanism. }Under a mathematical point of view, this recall of learned
patterns occurs since synaptic modifications during learning make
the stored patterns to become attractors of the underlying dynamics
of the system \cite{amitB,perettoB}. Then, if the input pattern
puts the network activity within the basin of attraction of a given
memory, the system can retrieve it.

The brain of mammals is able to recognize visual information or other
sensory patterns by means of this associative memory mechanism \cite{asociative0,asociative1}.
When information received through the senses reaches the brain sensory
areas, neurons become firing or silent according to some information
coding scheme \cite{Gutnisky2008}. It is well established, that
this information activates these neuronal areas in such a way that
when an excitatory neuron fires it induces the firing of neighboring
neurons. To achieve this, some modifications on the synaptic weights
has to be done any time neighboring neurons are active or inactive
during the processing of sensory information \cite{castro95,memory13}.
One of the most common used paradigms for learning is the so called
Hebbian learning rule \cite{hebb}, which is often summarized as
\emph{\textquotedbl{}Cells that fire together, wire together\textquotedbl{}.
}Attending to this rule, the group of neurons related, for instance,
to the green color and those to the smell of the field grass would
fire together, and their connections would become stronger so that
both feelings will activate correlated areas inside our brains \cite{senses94}.
Hebbian learning paradigm has been widely described to be present
in different biological systems including brain mammals as in the
traditional experiments of long-term potentation (LTP) \cite{ltprat1996}
but also in invertebrates as in the honeybee antennal lobe \cite{Galan2006}
(where it serves as an olfactory sensory memory). Those two pieces
of the same puzzle, namely a Hopfield like network -- which provides
the structure and the dynamic of a neural network -- and Hebbian learning
that is responsible of creating the optimal synaptic weights in order
to retrieve predefined stored patterns of activity, works properly
and they constitute the basics elements when trying to model and simulate
auto-associative tasks in the brain.

On the other hand, it has been long observed that in mammals cortex
there exists a balance between excitatory and inhibitory neurons \cite{Shubalanced2003,balance_scholar,balance1}
which seems important to regulate the activity in actual neural systems
\cite{Dani30082005}. Traditionally Hopfield-like models do not account
properly for this balance (although, as it is shown below, the Hebbian
learning rule implies for random patterns the same amount of excitation
and inhibition in these networks). In this work, we analyzed the implications
that result from including a biophysically motivated balance of excitation
and inhibition in autoassociative neural networks. We introduced this
balance by adding a new random term with probability $(1-c)$ to Hebbian
synaptic intensities -- which occurs then with probability $c$ --
drawn from a bimodal distribution. In order to mimic the experimental
findings, the excitatory mode in this distribution has a probability
which is four times larger than the inhibitory one but where its strength
is four times lower. The resulting synaptic intensities satisfy the
excitation/inhibition balance found in actual neural systems, and
due to the presence of the Hebbian term with probability $c$, the
model preserves the associative memory property for some regions of
the relevant parameters.

Additionally, our system presents new intriguing features at low temperatures,
even in the limit of the loading parameter {$\alpha=\lim_{N\rightarrow\infty}P/N\rightarrow0,$
with $P$ and $N$ being, respectively, the number of stored patterns
and the network size.} This includes the appearance at low temperatures
of new type of stable sates where the associative memory property
is lost, and which are embedded among the (also stable) traditional
memory states. This multistable phase is such that the low-noise non-memory
or \emph{frustrated} states have largest basin of attractions than
the traditional memory attractors. Moreover, for values of $c$ below
some given value, the memory attractors become metastable so the frustrated
states turn into the global attractor of the dynamics of the system.
Besides, the frustrated attractors are not correlated with memory
ones and are characterized by other type of order so that, depending
on initial conditions, these states can correspond to a high activity
or Up state, or to a low activity or Down state. Moreover, these frustrated
states can not be the same as the traditional spin-glass states appearing
in the Hopfield model at low temperature, since in this last case
these type of states appear only for $\alpha>0.$

Future extension of our study here could be useful to understand how
synapse heterogeneity can lead to the appearance, for instance, of
Up/Down states in the mammalian cortex not correlated with memory
attractors, and to explore the appearance of transitions among these
states -- as those observed during the sleep-wake transitions or anesthesia
\cite{Destexhe2007} -- when some attractor destabilizing mechanisms,
as for instance dynamic synapses \cite{torresNC,cortes06,mejias10updown}
or hyperpolarizing potassium slow currents \cite{benita12}, are
incorporated in the neural network model.

\section{Models and Methods}

Our starting point is a network of $N$ binary neurons whose possible
states $s_{i}=0,1;\,\forall i=1,\ldots,N$ represent neurons in a
silent or firing state, respectively. We then define a state evolution
for the network, where each neuron obeys the following probabilistic,
parallel and synchronous dynamics \cite{perettoB}: 
\begin{equation}
P[s_{i}(t+1)=1]=\frac{1}{2}\left\{ 1+\tanh\left[2\beta\left(h_{i}(\mathbf{s},t)-\theta_{i}\right)\right]\right\} \quad\forall i=1,\ldots,N,\label{eq:dinamica}
\end{equation}
where $h_{i}(\mathbf{s},t)$ is the local field or the total input
synaptic current arriving to neuron $i$, and which is defined as
\begin{equation}
h_{i}(\mathbf{s\mathrm{,t}})=\sum_{j\neq i}\omega_{ij}s_{j}(t)\varepsilon_{ij}.\label{eq:imput_current}
\end{equation}
Here $\varepsilon_{ij}$ is the connectivity matrix, that for simplicity
we shall assume here to be that of a fully connected network, i.e.
$\varepsilon_{ij}=1\forall i\neq j$ and $\epsilon_{ii}=0$ to avoid
self-connections. The variable $s_{j}(t)$ represents the current
sate of the $j^{th}$ presynaptic neuron and $\omega_{ij}$ is the
matrix of synaptic weights where each element specifies the strength
of the connection between the $i^{th}$ and $j^{th}$neurons. {We
denote also $\beta\equiv T^{-1}$ as the inverse of a temperature
parameter $T$ -- controlling the level of thermal noise in the system
-- such that $\beta\rightarrow\infty$ ($T=0$) implies a deterministic
dynamics whereas for $\beta=0$ ($T\rightarrow\infty$) the network
evolves fully randomly.}

\begin{figure}
\begin{centering}
\includegraphics[scale=0.9]{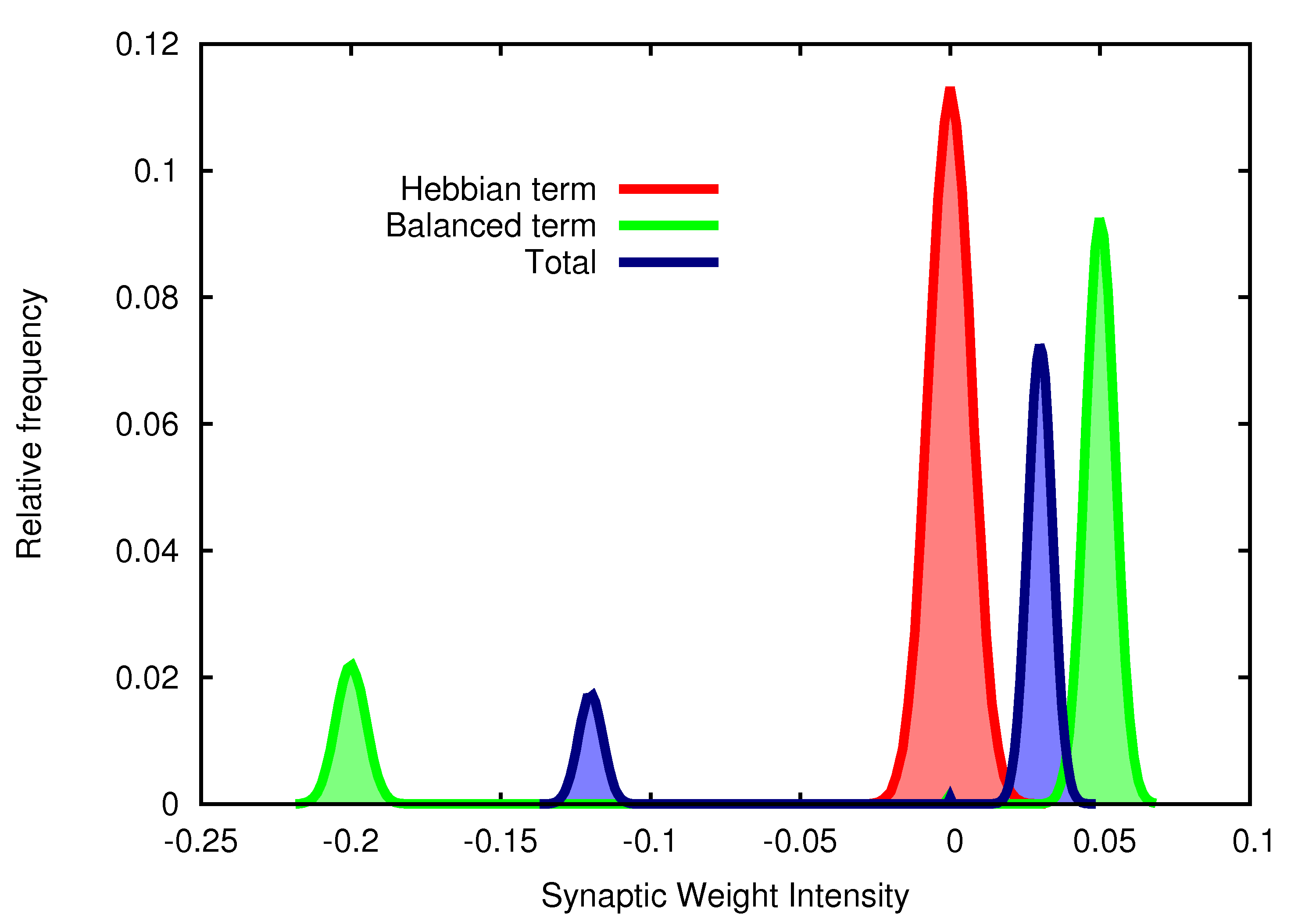} 
\par\end{centering}

\caption{Generation of balanced I/E synaptic weight distributions including
a Hebbian term. The red solid curve is the distribution of the synaptic
weights $\omega_{ij}^{H}$ considering only a Hebbian term, which
turns out to be {approximatively} a Gaussian distribution for a
large number $P$ of unbiased random patterns \cite{feller}. On
the other hand, the green solid curve represents a bimodal or balanced
E/I non-Hebbian Gaussian distribution. We build this balanced distribution
such that excitatory (positive) synaptic weights are more likely to
occur (four times more) than inhibitory (negative) synaptic weights,
and their strength (on average) is four times lower than the intensity
of inhibitory synapses. Using these two distributions, we can build
a balanced Hebbian distribution for a random variable which is the
sum of a Hebbian random synaptic weight and a non-Hebbian random balanced
synaptic weight according to (\ref{eq:pesos_sinapticos}). This final
distribution (blue curve) still preserves the balance I/E and other
features of the original non-Hebbian balanced distribution, whereas
it includes Hebbian features that allow for associative memory tasks.
Parameter values used to obtain different histograms were $N=1000,\,P=50,\,\kappa=1,\,\sigma=0.1$
and $c=0.4.$}
\label{balanced-distribution} 
\end{figure}

In our model, we assume that synaptic intensities or weights are given
as 
\begin{equation}
\omega_{ij}=c\omega_{ij}^{H}+\left(1-c\right)\omega_{ij}^{B}\label{eq:pesos_sinapticos}
\end{equation}
\[
\omega_{ii}=0
\]
with $0\leq c\leq1,$ and where we avoid self connections. The first
term of the right-hand side of (\ref{eq:pesos_sinapticos}) refers
to synaptic modifications due to learning. We assume that this term
is given by the standard Hebbian learning rule that memorizes $P$
prescribed patterns of neural activity in the form \cite{covariancerule,amitB}
\begin{equation}
\omega_{ij}^{H}=\frac{1}{a(1-a)N}\sum_{\mu}\left(\xi_{i}^{\mu}-a\right)\left(\xi_{j}^{\mu}-a\right).\label{eq:aprendizaje_hebb}
\end{equation}
Here, $\left\{ \xi_{i}^{\mu}=0,1\,;\,i=1,\ldots,N\right\} $ represents
the $P$ stored random patterns with probability distribution \cite{covariancerule,amitB}
\begin{equation}
p\left(\xi_{i}^{\mu}\right)=a\delta\left(\xi_{i}^{\mu}-1\right)+\left(1-a\right)\delta\left(\xi_{i}^{\mu}\right)\label{eq:patrones}
\end{equation}
where $a=\left\langle \xi_{i}^{\mu}\right\rangle $ is the mean level
of neural activity in the pattern and $\delta(x)$ is the Dirac delta
function. On the other hand, $\omega_{ij}^{B}$ in (\ref{eq:pesos_sinapticos})
accounts for a random contribution to the synaptic weights not directly
associated to learning, as for instance synaptogenesis, synapse remodeling
and homeostatic synaptic plasticity during life brain development
\cite{Turrigiano2004}, synaptic pruning during early stages of brain
development \cite{huttenlocher1987,Chechik1998,Xu2009}, etc. Here,
we consider $\omega_{ij}^{B}$ to be a random number drawn from the
Gaussian bimodal distribution 
\begin{equation}
p\left(\omega_{ij}^{B}\right)=\eta\mathcal{N}\left(\kappa\alpha,\sigma^{2}\right)+\left(1-\eta\right)\mathcal{N}\left(-4\kappa\alpha,\sigma^{2}\right)\label{eq:termino_balanceado}
\end{equation}
with a main synaptic mode being excitatory (positive) and the other
inhibitory (negative), where $0\leq\eta\leq1$ is a parameter controlling
the probability of each synaptic mode, and $\sigma^{2}$ is the variance
of each distribution mode. {Note that in this way $\omega_{ij}^{B}$
and $\omega_{ji}^{B}$ are random numbers not necessary equal but
following the same distribution.} Also in (\ref{eq:termino_balanceado})
$\alpha\equiv P/N$ is the loading parameter that measures the amount
of patterns stored in the network relative to its size, and $\kappa>0$
is a arbitrary constant that controls the synaptic strength separation
between the two modes. {For a large number $P$ of independent unbiased
random patterns and due to the ``Central Limit Theorem'', the Hebbian
variable $\omega_{ij}^{H}$ becomes approximatively Gaussian \cite{feller}.}
Moreover it can be analytically proof that with the choice (\ref{eq:termino_balanceado})
the model reproduces a balance between excitatory synapses $\left(\omega_{ij}>0\right)$
and inhibitory synapses $\left(\omega_{ij}<0\right)$ in the network,
and that the strength of inhibitory synapses is four times the strength
of the excitatory ones. More precisely, the resulting total synaptic
weights $\omega_{ij}$ in (\ref{eq:pesos_sinapticos}) are also bimodal
distributed as it is depicted in figure \ref{balanced-distribution}
preserving the balance introduced by $p(\omega_{ij}^{B}).$ 

{The parameter $\eta$ measures the relative abundance of excitatory
synapses compared with inhibitory ones. Although a complete study
of the behavior of the system can be done as a function of this parameter
(see next section), we consider in most of our analysis appropriate
values of $\eta$ which can mimic actual conditions. For instance,}
for $\eta=0.8$ one obtains a balance between excitation and inhibition
similar to that reported to be present in the cortex of human brain
\cite{Shubalanced2003,balance1,balance_scholar}. With this choice
there is a $80\%$ of excitatory synapses and a $20\%$ inhibitory
ones. The factor $\alpha$ in (\ref{eq:termino_balanceado}) has been
included to impede the total synaptic current $h_{i}=\sum_{j}\omega_{ij}s_{j}\varepsilon_{ij}$
to diverge in the thermodynamic limit $(N\rightarrow\infty)$ {(in
some situations of interest this choice has to be done carefully,
see for instance \cite{diver1,diver2} for a discussion concerning
this issue)}. Finally, we define the threshold term appearing in
equation (\ref{eq:dinamica}) as $\theta_{i}=\frac{1}{2}\sum_{j}\omega_{ij}$,
to recover in the limit of $c=1$ the standard Hopfield model {in
the $\{1,-1\}$ code (see for instance \cite{Amit198730})}.

We can calculate how well a stored pattern is retrieved by the network
dynamics by means of the overlap function defined as 
\begin{equation}
m^{\mu}(\mathbf{s})\equiv\frac{1}{Na(1-a)}\sum_{i}\left(\xi_{i}^{\mu}-a\right)\left(s_{i}-a\right).\label{eq:haming}
\end{equation}
Due to the symmetry pattern-antipattern inherent in the present model,
a given memory $\mu$ will be retrieved during the dynamics of the
system when $|m^{\mu}(\mathbf{s})|=1.$ We can also measure the activity
of the neural network, that is, how many neurons are firing together
at the same time, monitoring the order parameter 
\begin{equation}
m(\mathbf{s})\equiv\frac{1}{N}\sum_{j}\left(2s_{j}-1\right).\label{eq:firing rate}
\end{equation}
This measure can be easily related with the mean network activity
or mean firing rate $\nu(\mathbf{s})=\frac{1}{N}\sum_{i}s_{i}=\frac{m(s)+1}{2},$
in such a way that a solution with $m(\mathbf{s})=1$, corresponds
to a neuron population with a high mean firing rate ($\nu=1$) (Up
state) and $m(\mathbf{s})=-1$ corresponds with a silent neural population
($\nu=0$) (Down state).

{In the next section we will report the main results of our study
for the case of a single stored pattern ($P=1$) which allow for a
simple mean-field theoretical treatment due to the lack of effects
produced by the interference of a number of stored patterns scaling
with $N.$}

\section{Results}

\subsection{Mean-field analysis\label{sub:Mean-Field-Approach}}

{In order to develop a theoretical treatment of the model presented
above, we have to note that the total synaptic weights (\ref{eq:pesos_sinapticos})
are intrinsically asymmetric due to the balanced term $\omega_{ij}^{B}$,
so one can not use typical theoretical techniques from equilibrium
statistical mechanics to derive self-consistent equations for the
order parameters (see for instance \cite{perettoB}). However, since
our system involves a fully connected network, we still can derive
a standard mean-field description to find these self-consistent equations
that will allow to monitor the behavior of the system as relevant
control parameters change. We can start by computing the quenched
and canonical ensemble average of some magnitudes of interest in their
steady state using the thermodynamic limit $\left(N\rightarrow\infty\right)$.
For example, for a given realization of the quenched disorder $\{\xi_{i}^{\mu},w_{ij}^{B}\}$
and from equation (\ref{eq:haming}) the mean overlap function becomes}
\begin{equation}
m^{\mu}\equiv\langle m^{\mu}(\mathbf{s})\rangle=\frac{1}{Na(1-a)}\sum_{i}\left(\xi_{i}^{\mu}-a\right)\left(\left\langle s_{i}\right\rangle -a\right),
\end{equation}
where using (\ref{eq:dinamica}) one obtains 
\begin{equation}
\left\langle s_{i}\right\rangle =\frac{1}{2}\left[1+\tanh\left[2\beta\left(h_{i}-\theta_{i}\right)\right]\right].\label{smean}
\end{equation}
{For simplicity, in the following we will consider the case of a
single stored random pattern $(P=1)$ with equal probability of having
$1$ or $0$ for its elements, i.e. $p\left(\xi_{i}^{\mu}\right)=\frac{1}{2}\delta\left(\xi_{i}^{\mu}-1\right)+\frac{1}{2}\delta\left(\xi_{i}^{\mu}\right),$
which implies $a\equiv\langle\xi_{i}^{\mu}\rangle=\frac{1}{2}$, and
then is straightforward to obtain} 
\[
\begin{array}{c}
2\left(h_{i}-\theta_{i}\right)=\frac{c}{N}\sum_{j}\left(2\xi_{i}^{\mu}-1\right)\left(2\xi_{j}^{\mu}-1\right)\left(2\left\langle s_{i}\right\rangle -1\right)+\left(1-c\right)\sum_{j}\omega_{ij}^{B}\left(2\left\langle s_{i}\right\rangle -1\right)\\
\\
=c\epsilon_{i}^{\mu}m^{\mu}+\left(1-c\right)Nm\omega_{i}^{B},
\end{array}
\]
where $\epsilon_{i}^{\mu}\equiv(2\xi_{i}^{\mu}-1)$ and 
\begin{equation}
m\equiv\langle m(\mathbf{s})\rangle=\frac{1}{N}\sum_{j}\left(2\langle s_{j}\rangle-1\right).\label{eq:parametro_m}
\end{equation}
In the last, we also assumed the particular case that $\omega_{ij}^{B}=\omega_{i}^{B}\;\forall j$,
that is, the random contribution to all synapses toward a given postsynaptic
neuron $i$ is of the same type (either excitatory or inhibitory)
and has the same random value or strength. Note additionally that
$\epsilon_{i}^{\mu}$ is also a random variable with probability $p(\epsilon_{i}^{\mu})=\frac{1}{2}\delta(\epsilon_{i}^{\mu}-1)+\frac{1}{2}\delta(\epsilon_{i}^{\mu}+1).$
We then finally can obtain the fixed point equations 
\begin{equation}
\begin{array}{l}
m^{\mu}=\frac{1}{N}\sum_{i}\epsilon_{i}^{\mu}\tanh\beta[c\epsilon_{i}^{\mu}m^{\mu}+(1-c)Nm\omega_{i}^{B}]\\
\\
m=\frac{1}{N}\sum_{i}\tanh\beta[c\epsilon_{i}^{\mu}m^{\mu}+(1-c)Nm\omega_{i}^{B}].
\end{array}\label{eq:premean}
\end{equation}
{In the thermodynamic limit $N\rightarrow\infty,$ one can assume
self-averaging of both order parameters and, therefore, the sums $\frac{1}{N}\sum$
appearing in (\ref{eq:premean}) can be replaced by averages over
the joint distribution of quenched disorder $P(\epsilon,\omega^{B})=p(\epsilon)p(\omega^{B}),$
leading to } 
\begin{equation}
\begin{array}{l}
m^{\mu}=\langle\langle\epsilon_{i}^{\mu}\tanh\beta[c\epsilon_{i}^{\mu}m^{\mu}+(1-c)Nm\omega_{i}^{B}]\rangle\rangle\\
\\
m=\langle\langle\tanh\beta[c\epsilon_{i}^{\mu}m^{\mu}+(1-c)Nm\omega_{i}^{B}]\rangle\rangle
\end{array}\label{eq:222}
\end{equation}
where $\langle\langle\cdot\rangle\rangle$ means average over $P(\epsilon,\omega^{B})$.
Assuming moreover the case of small $\sigma$ ($\sigma\rightarrow0$),
the Gaussian distributions in (\ref{eq:termino_balanceado}) can be
approached to delta distributions, that is 
\[
p(\omega_{i}^{B})\approx\eta\delta(\omega_{i}^{B}-\kappa/N)+(1-\eta)\delta(\omega_{i}^{B}+4\kappa/N).
\]
{We have checked in simulations other possible non-zero but small
values of $\sigma$ and the main results presented in this work (see
below) are not dramatically affected by this choice as long as the
synaptic intensities are sharply distributed around the two main modes
of (\ref{eq:termino_balanceado}). Note moreover that for $P=1$ one
has $\alpha=1/N.$ Then, after computing the double angle averages,
the system (\ref{eq:222}) becomes} 
\begin{equation}
\begin{array}{l}
m^{\mu}=\frac{\eta}{2}\tanh\beta(cm^{\mu}+(1-c)\kappa m)+\frac{1-\eta}{2}\tanh\beta(cm^{\mu}-(1-c)4\kappa m)\\
\\
+\frac{\eta}{2}\tanh\beta(cm^{\mu}-(1-c)\kappa m)+\frac{1-\eta}{2}\tanh\beta(cm^{\mu}+(1-c)4\kappa m),\\
\\
m=\frac{\eta}{2}\tanh\beta(cm^{\mu}+(1-c)\kappa m)+\frac{1-\eta}{2}\tanh\beta(cm^{\mu}-(1-c)4\kappa m)\\
\\
-\frac{\eta}{2}\tanh\beta(cm^{\mu}-(1-c)\kappa m)-\frac{1-\eta}{2}\tanh\beta(cm^{\mu}+(1-c)4\kappa m),
\end{array}\label{eq:steady_state}
\end{equation}

which in a more compact form can be written as 
\[
\begin{array}{c}
m^{\mu}+m=\eta\tanh\beta(cm^{\mu}+(1-c)\kappa m)+(1-\eta)\tanh\beta(cm^{\mu}-(1-c)4\kappa m)\\
\\
m^{\mu}-m=\eta\tanh\beta(cm^{\mu}-(1-c)\kappa m)+(1-\eta)\tanh\beta(cm^{\mu}+(1-c)4\kappa m).
\end{array}
\]
One can solve numerically the steady-state equations (\ref{eq:steady_state})
and find fixed point solutions for $m^{\mu}$ and $m$ as a function
of the temperature parameter $T$ and for several values of the parameter
$c$ (see below). However, one can analytically study some limits
of interest. For example, let us consider the limiting case $c=1.$
In this situation, one recovers the mean field equation for the standard
Mattis states of the Hopfield model \cite{Amit198730} 
\[
\begin{array}{l}
m^{\mu}=\tanh\beta m^{\mu}\\
\\
m=0.
\end{array}
\]
The first equation of the last expression has the trivial solution
$m^{\mu}=0,$ which is the only one for large $T.$ For lower values
of $T,$ this equation can be developed for small $m^{\mu}$ given
a critical temperature $T_{cr}=1$ for the appearance of memory Mattis
states ($m^{\mu}\neq0$). On the other hand, for the limit $c=0$
one has 
\begin{equation}
\begin{array}{l}
m^{\mu}=0\\
\\
m=\eta\tanh\beta\kappa m-(1-\eta)\tanh4\beta\kappa m.
\end{array}\label{eq:c0}
\end{equation}
Then, for $c=0,$ there are not Mattis states (since $m^{\mu}=0$),
so memories states are not present. However, other type of network
activity states (not yet described in the literature) can emerge,
namely the non-trivial solutions with $m\neq0$ of the second equation
of (\ref{eq:c0}).

One easily can visualize that $m=0$ is a trivial solution of the
second equation of (\ref{eq:c0}), which is the only one present for
very large $T.$ On the other hand, for $T=0$, one has that the equation
for $m$ becomes $m=(2\eta-1)\mbox{sign}(m)$. Then, e.g, for the
case of $\eta=0.8$, this gives a nontrivial Up state with $m=0.6\,(i.e.,\,\nu=0.8),$
and a Down state with $m=-0.6\,(i.e.,\,\nu=0.2).$ The critical temperature
for the appearance of non-trivial solutions with $m\neq0$ can be
easily computed developing the second equation of (\ref{eq:c0}) for
small $m$ which gives 
\[
\overline{T}_{cr}=5\eta\kappa-4\kappa.
\]
This results in a real critical temperature only when $\eta>4/5$
(independently of $\kappa$), and non-trivial solutions ($m\neq0$)
appear below such $\overline{T}_{cr}$ as in a continuous second order
phase transition. Note that $\eta=0.8=4/5$ marks the limit for the
appearance of this second order phase transition. That is, for larger
values of the population of excitatory synapses, i.e, $\eta>0.8,$
the transition for the appearance of this non-trivial network activity
states is second order. On the other hand, for $\eta\le0.8$ there
is not such critical temperature, which implies a sharp first-order
phase transition for the appearance of these non-trivial network activity
states.

For $0<c<1,$ both type of solutions, that is Mattis memory states
and the Up and Down activity states not correlated with stored memories,
can also appear for different values of the temperature parameter
$T,$ and can coexist as we will illustrate in the next section. In
this more general scenario, one can derive also some analytic results
as follows. By a simple inspection of equations (\ref{eq:steady_state})
one can easily check that $m^{\mu}=0$ and $m=0$ is a trivial solution
of the these steady-state equations, and that such solution appears
for large $T,$ i. e. small $\beta,$ above some critical temperature.
Making a Taylor expansion of equations (\ref{eq:steady_state}) around
this trivial solution one finds: 
\[
\begin{array}{l}
m^{\mu}\approx\beta cm^{\mu}\\
\\
m\approx\beta(5\kappa\eta-4\kappa)(1-c)m.\\
\\
\end{array}
\]
This gives again a critical temperature $T_{cr}=c$ for the appearance
of memory Mattis states, and $\overline{T}_{cr}=(1-c)(5\kappa\eta-4\kappa)$
for the appearance of nontrivial states with $m\neq0.$ Note that
$T_{cr}\neq\overline{T}_{cr}$ which implies that states with $m^{\mu}\neq0$,
in general, do not correspond to states with $m\neq0,$ or that states
with $m^{\mu}=0$ are not usually associated with states with $m=0.$
Additionally, since the stored patterns have $\langle\xi_{i}^{\mu}\rangle=a=0.5,$
any time one reaches a steady state with $m^{\mu}\neq0$, it will
correspond to $m\approx0$ and viceversa. It is worth noting to say
that the cases $c=1$ and $c=0$ recover our previous findings above
in these limits of interest. In particular, that $\overline{T}_{cr}$
is a true critical temperature only for $\eta>0.8$. It is important
also to note that all results reported in next sections have been
obtained for $\eta=0.8,$ so there is not such critical temperature
$\overline{T}_{cr}.$ In this situation, the transition between states
with $m\neq0$ and states with $m=0$ occurs as in a first order transition
at a temperature that we denote as $T_{t}$ (see next section).

\subsection{Linear stability analysis and phase diagram}

The dynamics of our system can be easily derived from equation (\ref{eq:dinamica})
yielding to 
\[
\langle s_{i}\rangle_{t+1}=\frac{1}{2}\left[1+\tanh\left[2\beta\left(h_{i}(t)-\theta_{i}\right)\right]\right].
\]
After using the same analysis and approaches done in section \ref{sub:Mean-Field-Approach},
this dynamics results in the coupled iterative map: 
\begin{equation}
\begin{array}{l}
m^{\mu}(t+1)=\frac{\eta}{2}\tanh\beta(cm^{\mu}(t)+(1-c)\kappa m(t))+\frac{1-\eta}{2}\tanh\beta(cm^{\mu}(t)-(1-c)4\kappa m(t))\\
\\
+\frac{\eta}{2}\tanh\beta(cm^{\mu}(t)-(1-c)\kappa m(t))+\frac{1-\eta}{2}\tanh\beta(cm^{\mu}(t)+(1-c)4\kappa m(t)),\\
\\
m(t+1)=\frac{\eta}{2}\tanh\beta(cm^{\mu}(t)+(1-c)\kappa m(t))+\frac{1-\eta}{2}\tanh\beta(cm^{\mu}(t)-(1-c)4\kappa m(t))\\
\\
-\frac{\eta}{2}\tanh\beta(cm^{\mu}(t)-(1-c)\kappa m(t))-\frac{1-\eta}{2}\tanh\beta(cm^{\mu}(t)+(1-c)4\kappa m(t)),
\end{array}\label{eq:mean_field_map}
\end{equation}
which can be analyzed using standard techniques. 
\begin{figure}
\begin{centering}
\includegraphics[scale=0.5]{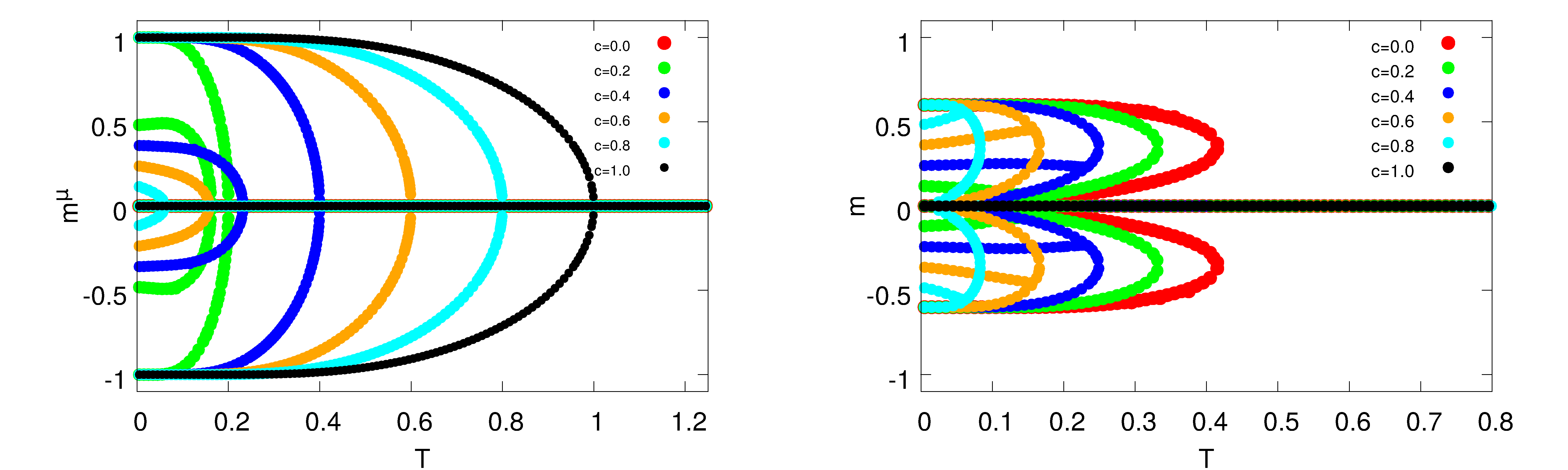} 
\par\end{centering}

\caption{Mean-field fixed point solutions for the overlap (left) and for the
mean network activity measured with $m$ (right) for different values
of $c$. Each colored curve represents the behavior as a function
of $T$ of all possible steady-state solutions (locally stable and
unstable) for a given $c$ value. Note the symmetry between positive
and negative solutions in both $m^{\mu}$(corresponding to pattern
and antipattern solutions) and $m$ (corresponding to Up and Down
activity states not correlated with a given stored pattern). Parameter
values were $\kappa=1,\,\eta=0.8$ and $\sigma=0.$}
\label{fig:Stability-of-the-todo} 
\end{figure}

\begin{figure}
\begin{centering}
\includegraphics[scale=0.55]{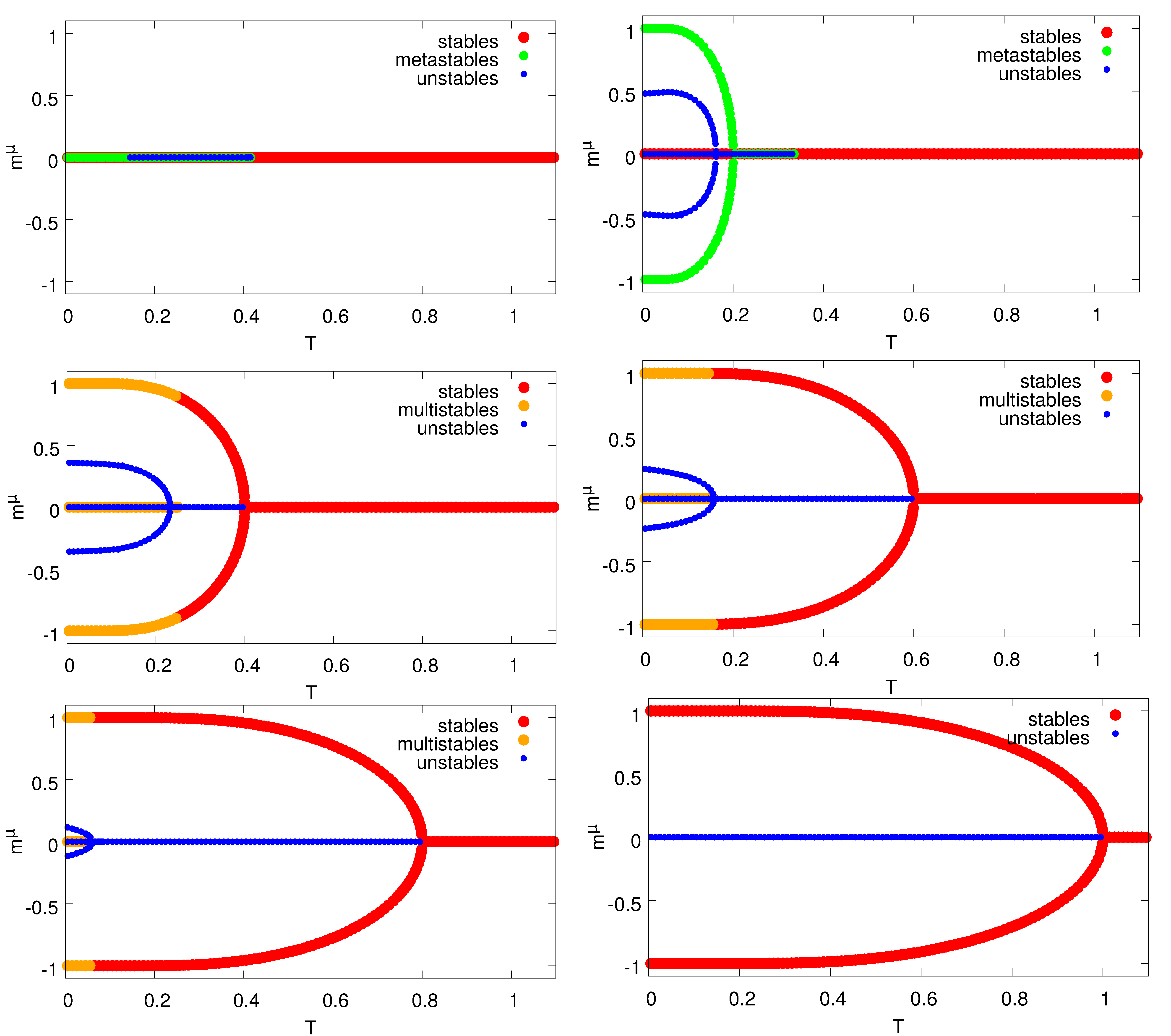} 
\par\end{centering}

\caption{Description of the local-stability of the overlap fixed point solutions.
Different steady solutions of the overlap function $m^{\mu}$ for
different values of $c$ are depicted. The values of $c$ used, starting
from top left to bottom right panels, are $c=0.0\:c=0.2\:c=0.4\:c=0.6\:c=0.8\:c=1.0$.
Bottom right figure shows the fixed-point solutions and their stability
for the standard Hopfield case $c=1$. Other panels corresponds to
the case in which balanced contribution to the synaptic weights is
present with probability $c.$ Parameter values have been taken as
in figure \ref{fig:Stability-of-the-todo}.}

\label{fig:Stability-of-overlap} 
\end{figure}

\begin{figure}
\begin{centering}
\includegraphics[scale=0.6]{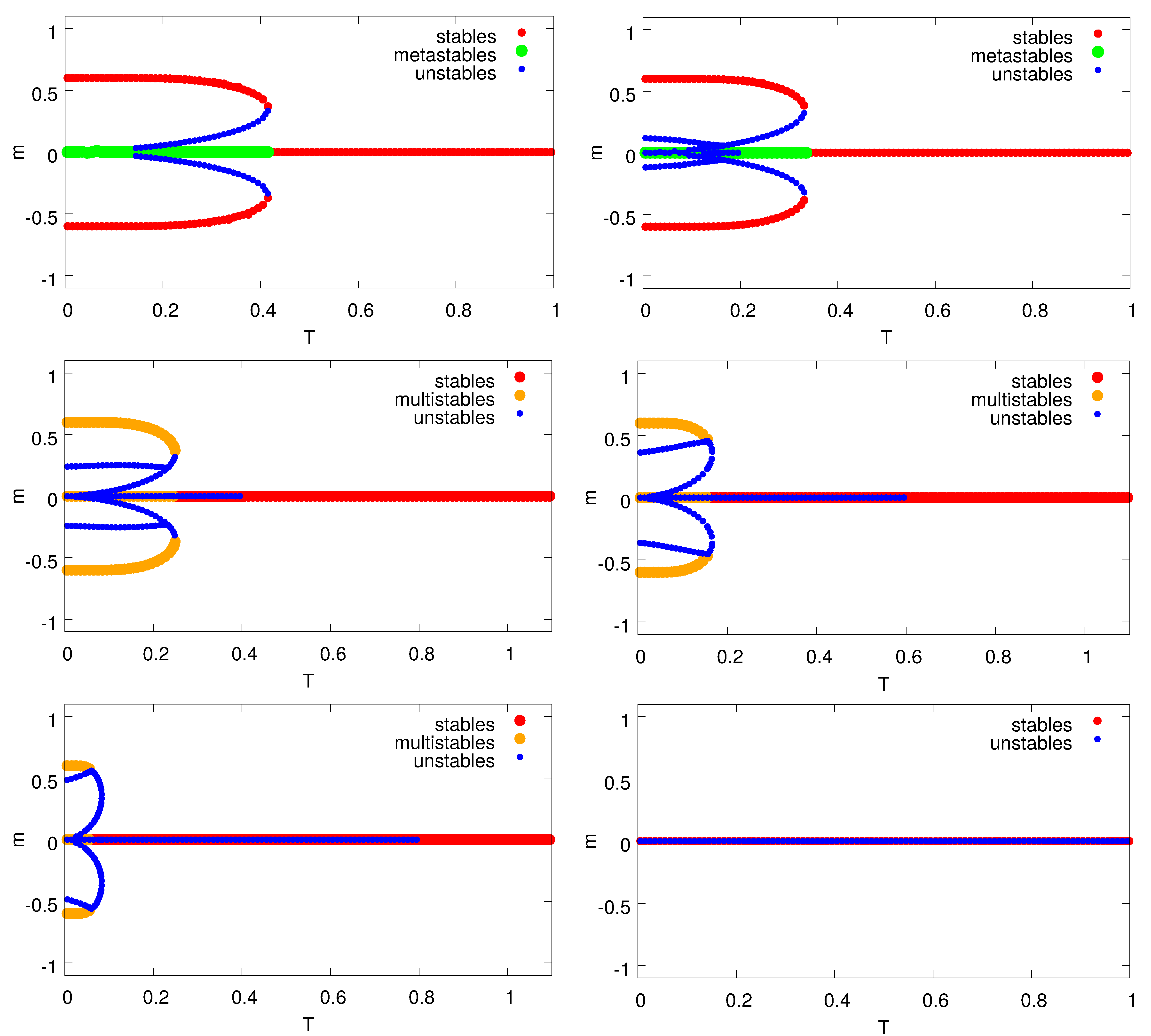} 
\par\end{centering}

\caption{Stability of mean network activity fixed-point solutions. Each panel
illustrates different steady-state solutions of the mean network activity
$m$ for different values of $c$. The values of $c$ used starting
from top-left to bottom-right panels are $\left[c=0.0\:c=0.2\:c=0.4\:c=0.6\:c=0.8\:c=1.0\right]$.
Bottom right panel shows the traditional Hopfield model solution where
the synaptic weights are computed according to Hebb rule $(c=1)$.
The figures clearly depict that in addition to a locally stable zero
$m$ solution, others locally stable non-zero fixed-point solutions
of $m$ emerge at very low temperatures below certain transition temperature
$T_{t},$ as in a first order phase transition. This fact results
in multistability or metastability in the system just below $T_{t}.$ }

\label{fig:Stability-of-firing} 
\end{figure}

Fixed point solutions of such map, as a function of the relevant parameters
$T$ and $c$ are plotted in figure \ref{fig:Stability-of-the-todo},
which illustrates the complex and rich attractor structure of our
system. However, not all of these fixed point solutions are stable
so in simulations of the system some of them will no be reached in
stationary conditions. To see the stability of these solutions and
the bifurcation structure of our system in terms of $T$ and $c,$
we can perform, e.g., a standard linear stability analysis of the
fixed point solutions. This implies that we have to diagonalize the
resulting Jacobian Matrix $\mathbf{A}=(A_{ij})$ of the previous iterative
map (\ref{eq:mean_field_map}), that is 
\[
\begin{array}{l}
A_{11}=\eta\beta c(\mathcal{B}_{+}^{\mu}+\mathcal{B}_{-}^{\mu})+(1-\eta)\beta c(\mathcal{C}_{+}^{\mu}+\mathcal{C}_{-}^{\mu})\\
\\
A_{12}=\eta\beta(1-c)\kappa(\mathcal{B}_{+}^{\mu}-\mathcal{B}_{-}^{\mu})+4(1-\eta)\beta(1-c)\kappa(\mathcal{C}_{+}^{\mu}-\mathcal{C}_{-}^{\mu})\\
\\
A_{21}=\eta\beta c(\mathcal{B}_{+}^{\mu}-\mathcal{B}_{-}^{\mu})-(1-\eta)\beta c(\mathcal{C}_{+}^{\mu}-\mathcal{C}_{-}^{\mu})\\
\\
A_{22}=\eta\beta(1-c)\kappa(\mathcal{B}_{+}^{\mu}+\mathcal{B}_{-}^{\mu})-4(1-\eta)\beta(1-c)\kappa(\mathcal{C}_{+}^{\mu}+\mathcal{C}_{-}^{\mu})\\
\\
\mathcal{B}_{\pm}^{\mu}=\frac{1}{2}[1-\tanh^{2}\beta(cm^{\mu}\pm(1-c)\kappa m)];\,\mathcal{C}_{\pm}^{\mu}=\frac{1}{2}[1-\tanh^{2}\beta(cm^{\mu}\pm(1-c)4\kappa m)].
\end{array}
\]
The local stability criterion for a given steady state solution, let
say ($(m^{\mu})^{*},m^{*}$), then is $|\lambda|_{max}=max_{i}|\lambda_{i}|<1,$
where $\lambda_{i}$ are the eigenvalues of $\mathbf{A}$ evaluated
in the solution ($(m^{\mu})^{*},m^{*}$). Using this criterion we
depict with different colors in figure \ref{fig:Stability-of-overlap}
and \ref{fig:Stability-of-firing}, the stability of the steady-state
solutions previously illustrated in figure \ref{fig:Stability-of-the-todo}.
Blue colors corresponds to unstable solutions, that is, solutions
such that $|\lambda|_{max}\geqslant1,$ whereas red, orange and green
solutions are locally stables, that is, for these solutions one has
$|\lambda|_{max}<1.$ Both figures show that below certain transition
temperature $T_{t}$ and for certain range of $c$ values, metastability
or multistability can emerge in the system (there are more than one
fixed-point solution which are locally stable for the same value of
temperature). Metastable solutions are depicted, for instance, in
top panels of figures \ref{fig:Stability-of-overlap} and \ref{fig:Stability-of-firing},
whereas multistability is shown in middle panels and left bottom panel
of these figures. More precisely, the stable memory attractors (red
solutions) -- that have $m^{\mu}\neq0$ and $m=0$ -- remain stable
or metastable (orange or green solutions in figure \ref{fig:Stability-of-overlap})
below $T_{t}$ and, additionally, a stable non-memory solution emerges
also below $T_{t}$. These correspond to the orange and red solutions
with $m^{\mu}=0$ and $m\neq0$ appearing for $T<T_{t}$ in figure
\ref{fig:Stability-of-overlap} (see also figure \ref{fig:Stability-of-firing}).

\begin{figure}
\begin{centering}
\includegraphics[scale=0.6]{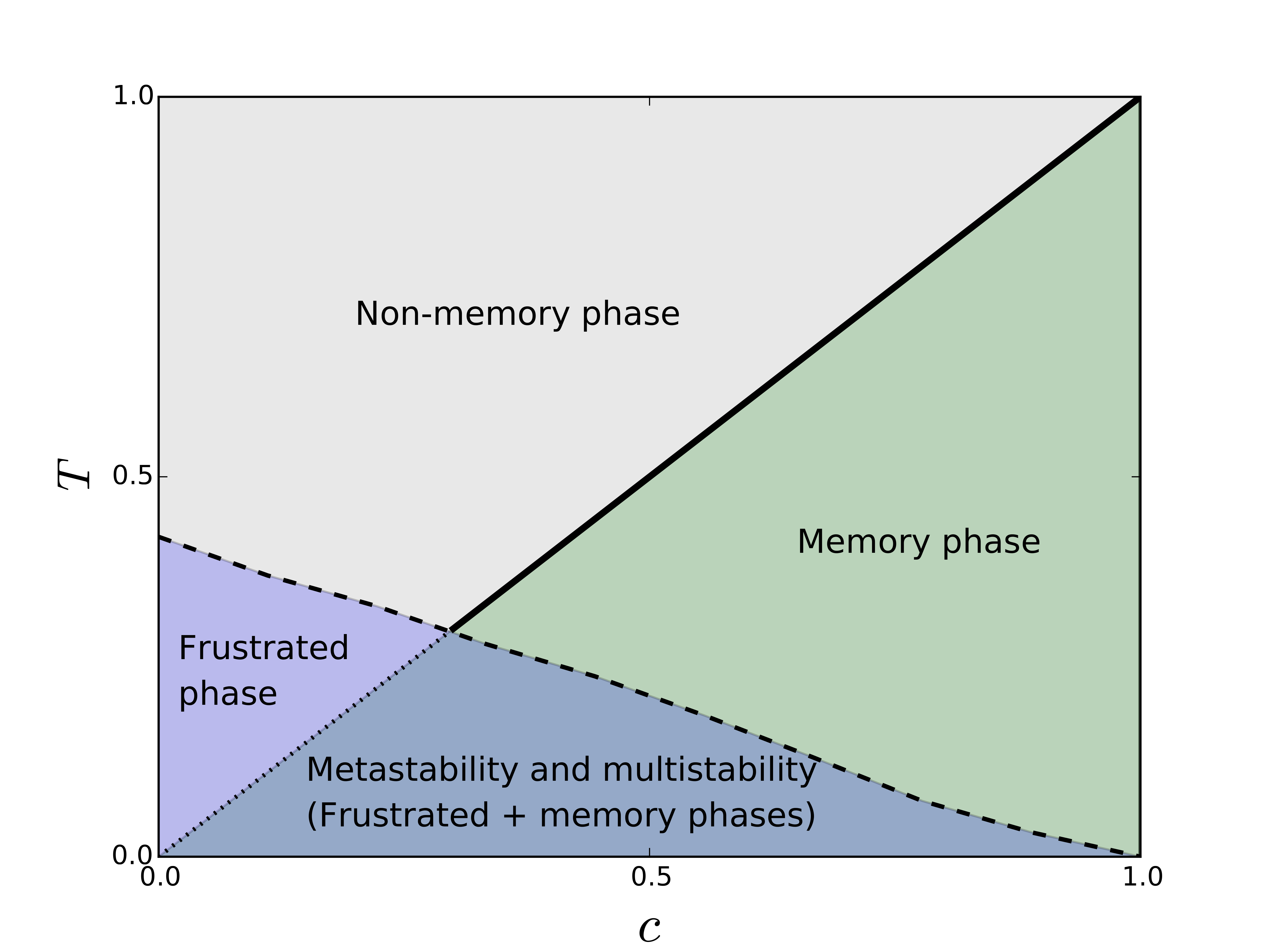} 
\par\end{centering}

\caption{Phase diagram depicting different phases in our system. For $T<T_{cr}=c$
(solid and dotted line) stable memory states appear as in a second
order phase transition. For $T<T_{t}$ (dashed line) frustrated states
emerge that are the only present in the system for $T>T_{cr}$ and
that can coexist with the memory states for $T<T_{cr}.$ Parameter
values have been taken as in figure \ref{fig:Stability-of-the-todo}.}

\label{fig:Phase-diagram} 
\end{figure}

Using the stability criterion explained above we have computed the
whole phase diagram ($T,c$) of our system for $\kappa=1$ and $\eta=0.8,$
which is depicted in figure \ref{fig:Phase-diagram}. This shows a
paramagnetic or non-memory phase (gray color) at high temperature
for the whole range of $c.$ As $T$ decreases a memory phase (dark
green color) emerges as in a second order phase transition at the
critical temperature $T_{cr}=c$, (solid and dotted lines). In addition,
also for low values of temperature a frustrated phase with non-memory
attractors emerges as in a sharp first-order transition below some
transition temperature $T_{t}=f(c)$ for all values of $c$ (dashed
line). For $T<T_{t}$ and $T<T_{cr}$ (dark blue region) frustrated
solutions coexist with the memory ones, in such a way that for $c\apprle0.3$
frustrated solutions are the global minima of the dynamics, being
memory solutions metastable, whereas for $c\apprge0.3$ multistability
among frustrated and memory states is present. Note that for $T<T_{t}$
and $T>T_{cr}$ , the frustrated phase is the only present in the
system (light blue color region). This type of non-memory attractors
differ from paramagnetic phase because of its high and low mean network
activities. As will be show in the next section, all these mean-field
results are confirmed by Monte Carlo simulations of our system.

\subsection{Monte Carlo simulations versus mean-field results}

\begin{figure}
\begin{centering}
\includegraphics[scale=0.7]{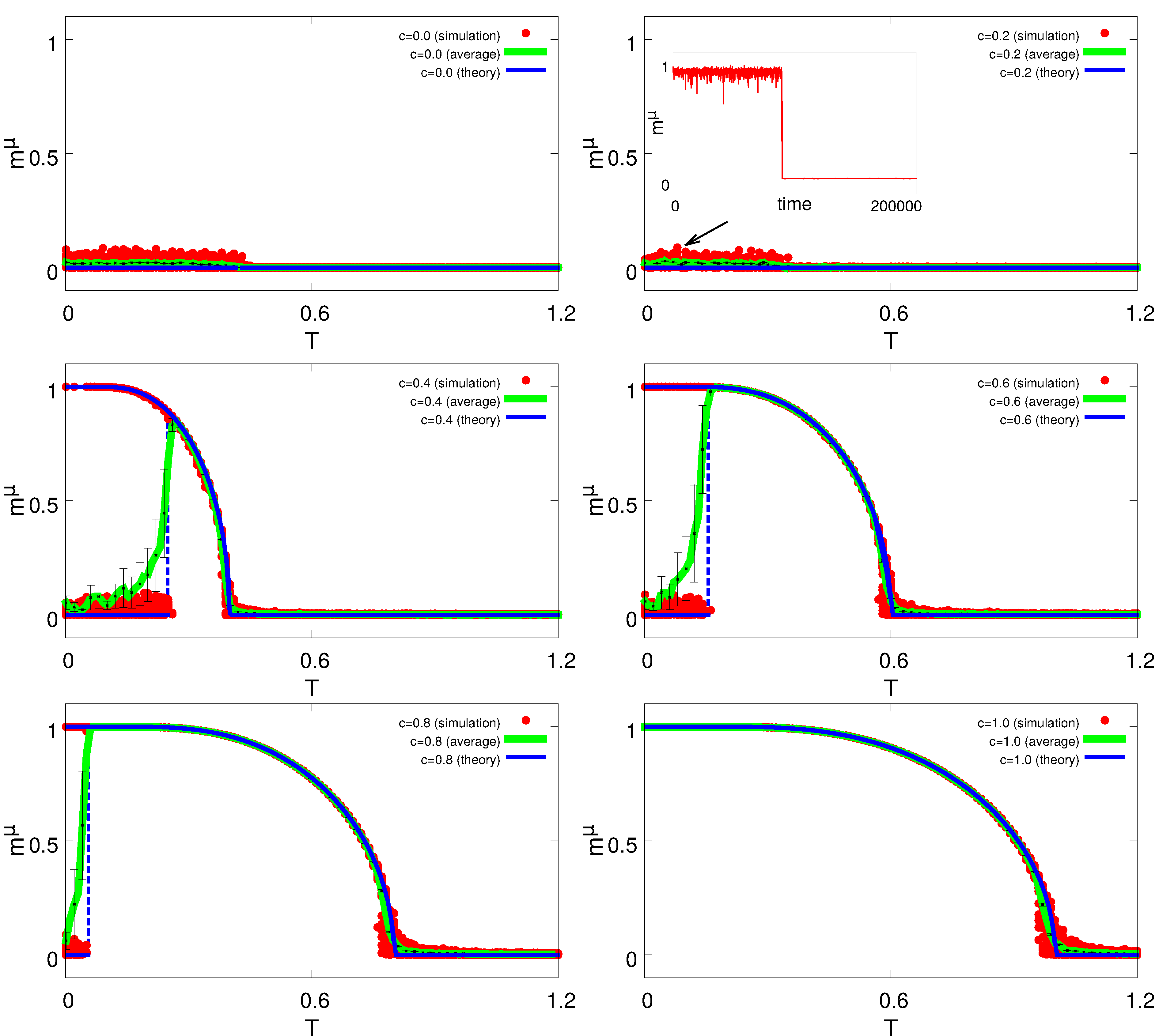} 
\par\end{centering}

\caption{Overlap vs Temperature. Each panel shows the behavior of the overlap
function $m^{\mu}$ for different values of $c=0,\,0.2,\,0.4,\,0.6,\,0.8,\,1$
(increasing from top-left to bottom-right). The top left figure $(c=0)$
is the situation where the Hebbian term in absent, while the bottom
right figure $(c=1)$ is the opposite situation, where only the Hebbian
term rules the synaptic strength. Blue solid line represent the mean-field
steady-state of the system. Red points corresponds to the final time-averaged
steady-state of the system for 50 realizations of the simulated systems
with $N=1600$ neurons, whereas the green solid line is the averaged
steady-state over these 50 realizations of the system (error bars
correspond to the standard deviations of the 50 data points).}

\label{montecarlo_overlap} 
\end{figure}

\begin{figure}
\begin{centering}
\includegraphics[scale=0.7]{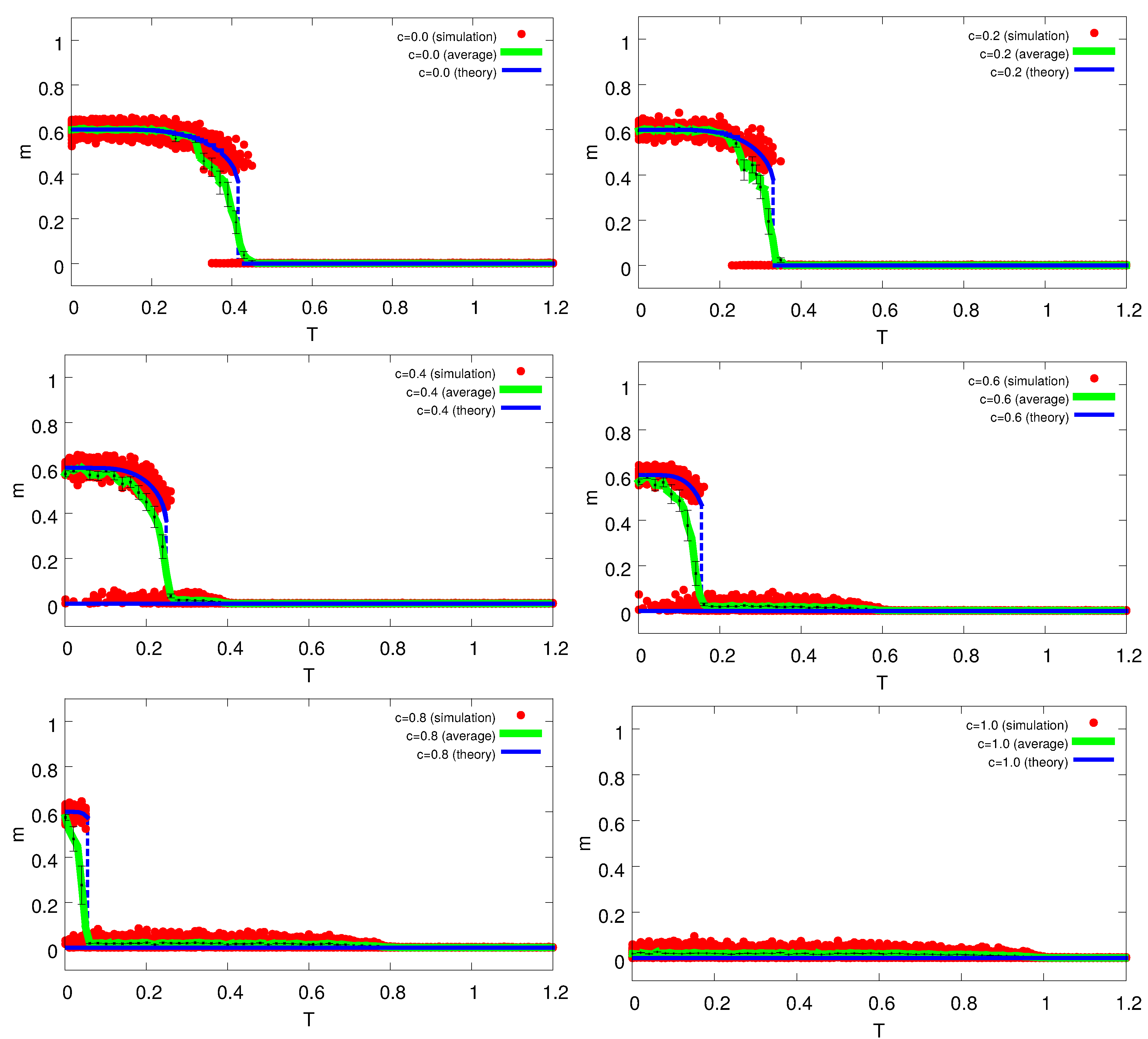} 
\par\end{centering}

\caption{Mean Network Activity vs Temperature for the same cases depicted in
figure \ref{montecarlo_overlap}. Each figure shows the behavior of
the mean network activity, as measured with the order parameter $m,$
for different values of $c$. Again, from top to bottom, the figure
depicts the steady state solutions that the system reaches both in
simulations (red data points) and within our mean field theory (blue
solid lines) as a function of $T,$ for increasing values of the parameter
$c.$}

\label{fig:Montecarlo_firing} 
\end{figure}

Monte Carlo simulations of our system are in agreement with the mean
field results described above for the whole range of the parameters
$T$ and $c$ we can consider, as the figures \ref{montecarlo_overlap}
and \ref{fig:Montecarlo_firing} clearly depict. In these figures
we only represent the positive solutions for $m^{\mu}$ and $m$ as
a function of $T$ and $c,$ but it is worth noting to say that due
to the symmetry $1,0$ for the state variables $s_{i}$ and stored
patterns $\xi_{i}^{\mu},$ exact symmetric negative solutions for
$m^{\mu}$ and $m$ are also present in the steady state of the system
not only in the mean-field results reported above but also in simulations.
In both figures, the blue solid lines correspond to the steady-states
obtained using the analytic treatment (mean-field approach) described
in the previous section, while the red data points are the steady-state
solutions obtained from the Monte Carlo simulations for a network
with $N=1600$ neurons. For a given value of $T$ and $c,$ these
points have been obtained for 50 realizations of the system, after
averaging time-series data points during a time window $\Delta t=500$
Monte Carlo steps (MCS) in the final steady-state. On the other hand,
the solid green curves has been calculated averaging the resulting
steady states (red points) over the 50 realizations of the simulated
system, where error bars represent the corresponding standard deviations.

The figure \ref{montecarlo_overlap} shows the behavior of the steady-state
overlap function $m^{\mu}$ as a function of the temperature $T$
for different values of the parameter $c$. It clearly illustrates
the appearance of frustrated states at low temperature where memory
is lost ($m^{\mu}=0$) for $T<T_{t}$ and where a new phase with Up
and Down activity states can emerge (see also \ref{fig:Montecarlo_firing}).
As in the mean-field results described in the previous section, in
simulations these frustrated states appears for any $c<1,$ and are
the only stable states for $c\lesssim0.3$ (see for instance the two
top panels in figure \ref{montecarlo_overlap} and the inset in the
top-right panel). The bottom-right panel of figure \ref{montecarlo_overlap}
corresponds to the traditional behavior of the Hopfield model ($c=1)$,
where there is no contribution of the synaptic balanced term. In this
case, one observes the expected memory retrieval curve as a function
of $T$, where memory recall becomes less effective as $T$ increases
below the critical temperature $T_{cr}=1$. As the parameter $c$
decreases, the term due to the balanced synaptic distribution becomes
more important and the critical temperature $T_{cr}$ goes down --
in fact as $T_{cr}=c$ -- since these synaptic balanced terms act
like a noise affecting the associative memory property. The range
of $T$ at which the associative memory emerges is also narrower as
$c$ decreases until it disappears for $c\lesssim0.3.$ In this situation,
in fact, memory attractors become metastable and the final steady-state
is a frustrated one (see inset in the top-right panel). Thus, e.g.,
for the other limit $c=0$, the associative memory property is completely
lost -- even metastable memory attractors are not present -- since
there is no contribution of the distribution governed by the Hebbian
rule.

On the other hand, figure \ref{montecarlo_overlap} also depicts that
the averaged steady-state positive overlap over different system realizations
(green solid line) has a reentrant behavior as a function of $T$
for $0.3<c<1.$ More precisely, for large $T$ the system falls into
the paramagnetic or non-memory phase so this averaged overlap is zero.
For values of $T<T_{cr},$ the averaged overlap follows the memory
recall curve that tries to saturate to its maximum value $1$ as $T$
decreases. This occurs until $T<T_{t}$ where the averaged overlap
again goes toward a small value near zero. This is due to the existence
of many frustrated states (with $m^{\mu}\approx0$) compared with
the number of steady-states that reach the stored pattern (with $m^{\mu}\approx1$).
Moreover, the number of realizations that put the system in the basin
of attraction of the memory state decreases as $T$ is lowered which
is an indication that the basin of attraction of the memory states
diminishes below $T_{t}$ when $T\rightarrow0.$ As soon as the relevance
of the balanced term is more important this quasi-reentrant transition
becomes more prominent (see the left-middle panel of figure \ref{montecarlo_overlap}
for $c=0.4$).

The above mean-field and simulation findings can be easily understand
since we are introducing a random contribution, with probability $(1-c),$
to the synaptic weights which does not depend on the stored patterns.
It is expected, therefore, that memory retrieval capacity of the system
decreases as $c\rightarrow0.$ The reason is that the mechanism involved
to achieve the recall of memories in auto-associative networks is
due to the correlation between the stored patterns $\xi_{i}^{\mu}$
and the current network activity $s_{i}$ state, and the information
content of memories is only present in the Hebbian term. Memory recall
capacity then decreases in a way that the critical temperature for
the appearance of Mattis states gets smaller as the random disorder
introduced by the bimodal contribution increases. This is clearly
depicted in the left-top panels of figures \ref{fig:Stability-of-overlap}
and \ref{montecarlo_overlap}, where it can be observed how the total
absence of Hebbian contribution in the learning rule, turns into memory
frustration due to the total lack of correlation between the patterns
and the synaptic weights.

As described above, simulations, in agreement with mean-field results,
shows the existence of a new intriguing region at low temperatures
(for $c<1$ below $T_{t}$) where the memory is frustrated (that is,
$m^{\text{\ensuremath{\mu}}}=0$) and where other type of order is
present (with $m\neq0$) as it is depicted in figures \ref{fig:Stability-of-firing}
and \ref{fig:Montecarlo_firing}. Note that in the memory phase one
has $m=0$ since we have chosen the learned pattern such that $\langle\xi_{i}^{\mu}\rangle=1/2$.
This new frustrated phase emerges sharply lowering the $T$ from the
stable memory phase as in a first order phase transition. One could
interpret this transition as a re-entrant phase transition from a
memory phase to a non-memory phase at low temperatures as in traditional
equilibrium and non-equilibrium spin-glass models \cite{ed_anderson1975,SK1975,torresprb98,marroB},
which in our case is of first order type. An important difference
with respect to previous studies, as in the traditional (equilibrium)
Hopfield model, is that the SG frustrated behavior is achieved for
infinitely large number of stored patterns, i.e. $\alpha>0$. Our
results here shows that frustrated states can be achieved also in
the case of $\alpha\rightarrow0$ for our model, when one assumed
a balance condition for synaptic weights and certain homogeneity among
the synaptic intensities of the afferents for a given postsynaptic
neuron.

Also in agreement with mean-field results above is the finding in
simulations that in this new phase, $m$ increases (decreases in the
case of the negative solution) as $T$ decreases until a value of
$m\approx0.6$ (or $-0.6$ in the case of the negative solution) at
$T=0.$ Moreover, simulations depict that this \emph{frustrated} phase
gets wider in the temperature range for $c\rightarrow0$ and, therefore
$T_{t}$ rises from zero as $c$ decreases reaching the value of $T_{t}\simeq0.42$
at $c=0$ -- see also the dashed line in figure \ref{fig:Phase-diagram}.
In terms of the mean firing rate $\nu,$ as we already mentioned,
the positive solution for $m$ corresponds with an Up state that has
a larger mean network activity (around $\nu\approx0.8$) which indicates
a state with a large population of firing neurons, whereas the symmetric
negative solution corresponds to a Down or silent state with very
low network activity (around $\nu\approx0.2$). On the other hand,
it is worth noting that memory states corresponds to a state with
average network activity around $0.5,$ i.e., a steady state with
approximately the same number of firing and silent neurons.

\begin{figure}
\begin{centering}
\hspace*{-1cm}\includegraphics[scale=0.18]{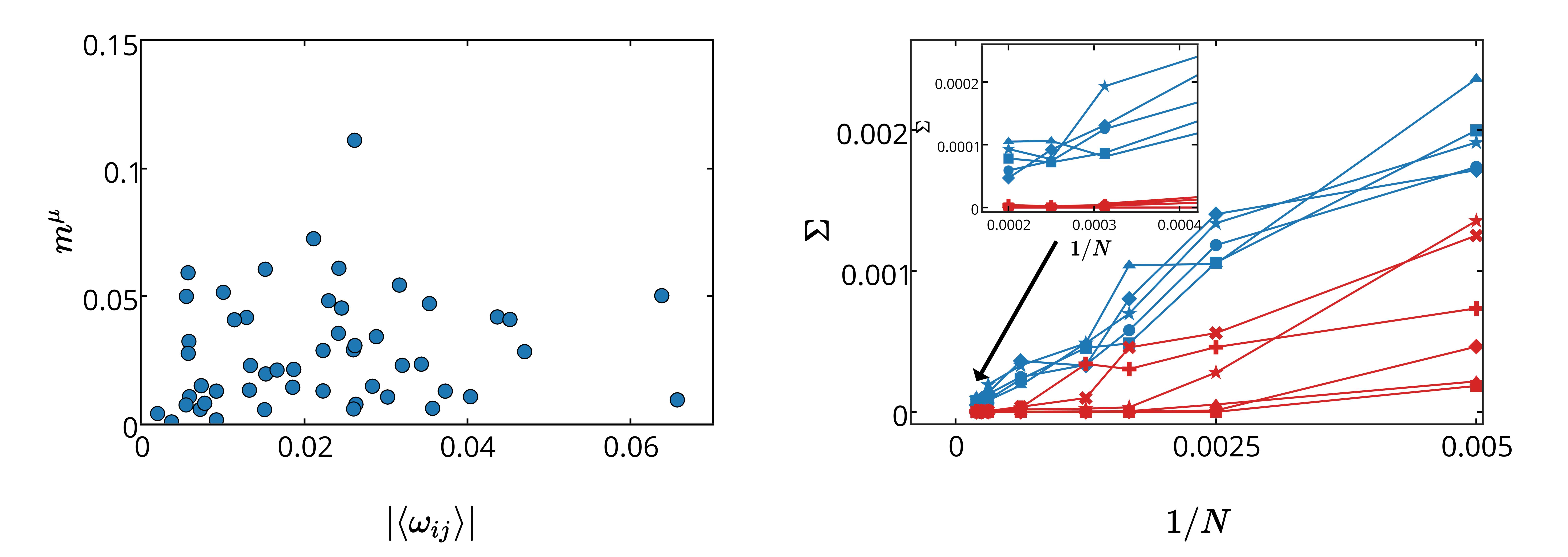} 
\par\end{centering}

\caption{Frustrated vs memory state fluctuations. Left panel depicts differences
in the final frustrated state (measured in terms of the overlap $m^{\mu}$)
over different realizations of the synaptic intensities, where $|\langle\omega_{ij}\rangle|$
is the absolute value of the mean synaptic strength of a given realization
of synaptic intensities through the whole network. Right panel depicts
the total size of the fluctuations of the steady state as a function
of the network size and different values for the temperature (different
symbols) for both, frustrated states (blue solid lines) and for memory
states (red solid lines). The inset shows a zoom of the large $N$
behavior of different lines which illustrates the saturation of the
size of the fluctuations for the frustrated states. Temperature values
considered were $T=10^{-3}(\CIRCLE),\,0.05(\blacksquare),\,0.1(\blacklozenge),\,0.15(\blacktriangle),\,0.2\,(\bigstar),$$0.25$(\ding{58}),$0.3$(\ding{54}).}

\label{sigma_compara} 
\end{figure}

In order to describe how frustration arises in these low-temperature
non-memory states, first we have checked if the steady-state averaged
overlap depends of the initial conditions. For a given realization
``$k$'' of synaptic intensities, namely $\mathbf{\omega}^{k}=\left\{ \omega_{ij}^{k}\right\} $,
and given values of the parameters $T$ and $c,$ we found that overlap
mean-value of the frustrated state remains the same for different
simulation initial conditions (data not shown). This indicates that
the system has only a minimum attractor associated to this frustrate
state. However, when one chooses different realizations of the synaptic
\emph{disorder} one finds different values of the overlap mean value
in the frustrated state as it is depicted in figure \ref{sigma_compara}Left.
This is an indication that the particular choice of the configuration
of synapses determines the features of the reached frustrated state
as in traditional spin-glass states.\textcolor{green}{{} }Due to
the homeostatic balance (more excitatory synapses, but stronger inhibitory
synapses) the synaptic intensities obey a constraint of the type,
$\langle\omega_{ij}\rangle\equiv\frac{1}{N^{2}}\sum_{i,j}\omega_{ij}=0$.
However, in our simulations $\langle\omega_{ij}\rangle$ is not exactly
zero consequence of the network finite-size, and the variability found
in the steady-state overlap mean-value of the frustrated solutions
is not correlated with the dispersion of $\langle\omega_{ij}\rangle$
from zero as figure \ref{sigma_compara}Left clearly depicts. This
demonstrates that dispersion of reached frustrated states has not
its origin in this finite-size effect and only is consequence of the
intrinsic synaptic disorder introduced by the balanced term.

Secondly, we have computed the size of fluctuations over different
realizations of synaptic disorder, for both memory and frustrated
states in the multistable region, for $c=0.4$ and different temperature
values, as it is shown in figure \ref{sigma_compara}Right. More precisely,
we computed

\[
\begin{array}{l}
{\displaystyle \Sigma^2=\frac{1}{N_{r}}\sum_{k=1}^{N_{r}}\left(m_{k}^{\mu}-\overline{m^{\mu}}\right)^{2}}\end{array}
\]

where $N_{r}$ is the number of realizations, $m_{k}^{\mu}$ is the
time-averaged overlap in the steady-state for a given synaptic disorder
realization ``$k$'', $\overline{m^{\mu}}\equiv\frac{1}{N_{r}}\sum_{k=1}^{N_{r}}m_{k}^{\mu}$
is the average of $m_{k}^{\mu}$ over different realizations of the
synaptic disorder, and $\Sigma^2$ is the variance
of $m_{k}^{\mu}$ also over different realizations of the synaptic
disorder. This figure illustrates how fluctuations in the frustrate
state are less sensible to temperature and remains even for very low
temperature. The figure also depicts that $\text{\ensuremath{\Sigma}}$
mainly depends on network size. Moreover, there is a tendency to saturate
the intensity of fluctuations toward a non-zero value for very large
$N$ (see blue curves in figure \ref{sigma_compara}Right and in its
inset). In the case of memory states, however, fluctuations intensity
decreases quickly with temperature until a zero value (see red curves
in figure \ref{sigma_compara}Right and in the inset). These findings
indicate that the frustrated state fluctuations might also appear
also in absence of thermal noise and in the thermodynamic limit and,
therefore, are only due to the synaptic heterogeneity introduced by
the balanced synaptic term.

\section{Conclusion}

In traditional autoassociative neural networks, as the Hopfield model,
the associative memory property is achieved by storing in the synaptic
weights the information to be learned by means of a Hebbian learning
prescription. In these models, each synaptic weight connects two different
neurons and its strength reflect the correlation between the activity
of those neurons during the activation of the learned pattern configuration.
It can be easily demonstrated that when the number of stored patterns
is very large, and these patterns are random and unbiased, the probability
to have a given synaptic strength value becomes Gaussian centered
around zero. This fact implies that in these models one has the same
probability to have excitatory and inhibitory synapses, contrary with
what is observed in nature, and that excitatory (or inhibitory) synapses
with a large strength are rare, that is, almost all synapses have,
in absolute value, a small value around zero.

In the present study, we have introduced a more realistic balanced
synaptic term -- with probability ($1-c$) -- in addition to the standard
Hebbian one (which then occurs with probability $c$). The resulting
probability distribution for the synaptic weights $\omega_{ij}$ has
more actual features, including an appropriate balance 4:1 between
excitation and inhibition, and large synaptic strengths of the inhibitory
synapses compared with the excitatory ones, as it has been described
in actual neural systems \cite{balance1,balance_scholar}. In addition,
we consider that the introduced balanced term does not include information
nor correlation with the stored patterns, so it acts as a ``balanced
synaptic noise'' against the memory order induced by the Hebbian
term.

As a consequence, the resulting model has new intriguing emerging
phenomena not observed in the standard Hopfield model. For instance,
for any value of $c<1,$ at low temperatures, a regime of memory frustration
arises as in a first order phase transition below some transition
temperature $T_{t}$. This kind of frustration resembles to that of
a spin glass (SG) state, where due to the randomness of the $\omega_{ij}$
and the absence of significant thermal fluctuations, the network state
can not retrieve any pattern and memory vanishes. However, the appearance
of SG like solutions in the standard Hopfield model is restricted
to the pattern saturation limit (that is $P\rightarrow\infty$ in
the thermodynamic limit $N\rightarrow\infty$) which is not the case
in our study, since frustrated states appears even for $\alpha=P/N=0$
($P$ finite in the thermodynamic limit). {It is worth noting to
say here that SG solutions can appear also in Hopfield like models
for $\alpha\rightarrow0$ when the number of stored patterns scale
with $N$ as $P\sim N^{k}$ with $k<1$ \cite{coolen1,coolen2} (which
is not the case either in our system)}.

At large temperature, the system undergoes a standard second-order
phase transition from the ferromagnetic (or memory) solutions to paramagnetic
(no-memory) ones at a critical temperature $T_{cr}=c$. This means
that the non-memory phase appears at lower temperatures as $c\rightarrow0$.
That is, the importance of the ``balanced synaptic noise'' diminishes
the memories attractor stability, so that thermal agitation needs
lower temperatures to destroy the memory and enter the system in a
paramagnetic phase.

Our main findings in this work observed by means of the Monte Carlo
simulations can be also be understand within a standard mean-field
approach of the model. In fact, a simple linear stability analysis
of the resulting mean-field dynamical equations demonstrated the existence
of a locally stable non-memory phase at large temperature above $T_{cr},$
a locally stable memory phase below this critical temperature and
multistability between memory and frustrated states at low temperature
below $T_{t}$ and intermediate values of $c$. On the other hand,
for low $c$, our analysis depicts that this multistable region transform
in another where the stable memory attractor becomes metastable in
such a way that the frustrated state becomes the global attractor
of the dynamics.

{Although we only present results concerning the case of $\alpha\rightarrow0,$
possible generalizations of the present study for $\alpha\neq0$ could
be thought using replica-like formalisms as in \cite{guerra1,guerra2}.
In this case, an important point to address is the relation between
the frustrated states we found in our work and different type of spin-glass
solutions emerging from the interference among many stored patterns
\cite{guerra3}. Also interesting is the possible relation of the
frustrated states we found in our system with the so called elusive
spin-glass states recently reported in \cite{elusiveSG}. These states
correspond only to a true spin-glass phase when the system is prepared
in an out of equilibrium state. In fact, our frustrated states only
appear for $c<1$ that correspond to the case of asymmetric weights
and therefore no Hamiltonian description is possible.}

Other important point in the present study is that the observed frustrated
states appearing at low temperature are associated with two different
levels of non-zero neuronal activity which is not correlated to that
of the stored patterns (since when the patterns are retrieved the
network activity becomes almost zero). These states\textbf{, }with
relative high and low activity, could be associated to the well known
``Up'' and ``Down'' states previously described in simple models
of interacting E/I neuron populations \cite{wilsoncowan72} and widely
observed in the cortical activity of the mammals \cite{sanchezvives00,Timofeev01122000}.
Our theoretical framework in this work could be easily extended to
include some destabilizing mechanisms of these states -- as for instance
including features of the hyperpolarizing potassium slow currents
or dynamic synapses -- which can induce transitions among then similar
to those observed in actual neural systems \cite{torresNC,tsodyks06,cortes07,mejias10updown,benita12}.

{Finally, some of the results and conclusions reported in the present
work could be used to design new paradigms of artificial neural networks
with possible applications in many fields such as robotics, artificial
intelligence, optimization methods, etc (see for instance \cite{booktonto}
and references therein).}

\section{Acknowledgments}

The present work has been done under project FIS2013-43201-P which
is funded by the Spanish Ministry of Economy and Competitiveness (MINECO)
and by the European Regional Development's Founds (FEDER). The authors
also thank J. Marro, P. L. Garrido and P. I. Hurtado for valuable
comments and suggestions.

\bibliographystyle{unsrt}
\bibliography{bibtotalnc}

\end{document}